%% file: chicJToLLo_Draft_New.tex
\documentclass[reprint,amsmath,amssymb,aps]{revtex4-1}
\usepackage{bm}
\usepackage{dcolumn}
\usepackage{lineno}
\usepackage{graphicx}
\usepackage{subfigure}
\usepackage{epstopdf}
\usepackage{float}
\usepackage{threeparttable}
\usepackage{overpic}
\usepackage{subfigure}
\usepackage{float}
\usepackage{cmap} 
\usepackage{mathtext}
\usepackage{mathrsfs}
\usepackage{multirow}
\usepackage{xcolor}
\definecolor{aa}{RGB}{0,0,139}
\usepackage[bookmarksnumbered=true,colorlinks,urlcolor=aa,linkcolor=blue,anchorcolor=aa,citecolor=aa]{hyperref}

\include{def-com}

\begin{document}
\title{Study of the decays  $\chi_{cJ}\to\Lambda\bar{\Lambda}\omega$}
\newpage
\input{authorlist_2023-12-11}

\date{\today}

\begin{abstract}
Using $(27.12\pm 0.14)\times10^{8}$  $\psi(3686)$ events collected with the BESIII detector, we present the first observation of the decays $\chi_{cJ}\to\Lambda\bar{\Lambda}\omega$, where $J=0, 1, 2$, with statistical significances of $11.7 \sigma, 11.2 \sigma$, and $11.8 \sigma$.  The branching fractions of these decays are determined to be $\Br(\chi_{c0}\to\Lambda\bar{\Lambda}\omega)=({2.37 \pm 0.22 \pm 0.23}) \times 10^{-4}$, $\Br(\chi_{c1}\to\Lambda\bar{\Lambda}\omega)=({1.01 \pm 0.10 \pm 0.11}) \times 10^{-4}$, and $\Br(\chi_{c2}\to\Lambda\bar{\Lambda}\omega)=({1.40 \pm 0.13 \pm 0.17}) \times 10^{-4}$, where the first uncertainties are statistical  and the second are systematic. We observe no clear intermediate structures.
\end{abstract}

\maketitle

\section{Introduction}\label{sec:introduction}

Decays of charmonium states offer insight into the behavior of confinement in Quantum Chromodynamics (QCD)~\cite{Brambilla:2010cs}.  Charmonium is unique in this regard since the charm quark mass scale resides between the perturbative and non-perturbative regions of QCD. To date, only a limited number of studies have been performed on the decays $\chi_{cJ}\to B\bar BM$ (where $B$ represents a baryon and $M$ denotes a meson), such as the decay $\chi_{cJ}\to\Lambda\bar{\Lambda}\eta$~\cite{zyj}. Hence, further studies are still highly desirable to improve our understanding of the properties of the $\chi_{cJ}$ states and the dynamics of their decays.

In previous studies of charmonium and bottomonium decays, several unanticipated enhancements with respect to phase space MC have been observed near the mass threshold of baryon anti-baryon pairs~\cite{xuwei, enhance_LLP, enhance_Belle}. Theoretical models, such as the quark-pair creation model, the one-boson exchange model, the $^{3}{P}_{0}$ model, and the Bonn meson-exchange model, have been used to interpret these enhancements~\cite{enhancement1, enhancement2}. Searching for $B\bar B$ mass  threshold enhancements in $\chi_{cJ} \to B\bar{B} M$ decays will improve our understanding of the underlying dynamics of charmonium decays. At the same time, we can search for potential excited baryon states in the $BM$ and $\bar{B}M$ invariant mass spectra, and search for new structures in the $B\bar{B}$ invariant mass spectrum. 

In addition, BESIII previously reported evidence for an excited $\Lambda$ state in the decay $\psi(3686) \to \Lambda \bar{\Lambda} \omega$~\cite{hhz}. Thus it is natural to extend the previous work to search for potential $\Lambda$ excited states in $\chi_{cJ}$ decays. The fact that the $\chi_{cJ}$ mesons have different spin-parity than the $\psi(3686)$ offers additional opportunities to investigate a $\Lambda\bar \Lambda$ mass threshold enhancement and possible excited states of the $\Lambda$. In this paper, we report the first observation of $\chi_{cJ}\to\Lambda\bar{\Lambda}\omega$, and search for a $\Lambda\bar \Lambda$ mass threshold enhancement and possible excited states of the $\Lambda$.  This is  based on $(27.12\pm 0.14)\times 10^{8}$ $\psi(3686)$ events~\cite{Ablikim:2017wyh} collected with the BESIII detector. 

\section{BESIII Detector and Monte Carlo Simulation}\label{sec:simu}\vspace{-0.3cm}

The BESIII detector~\cite{Ablikim:2009aa} records symmetric $e^+e^-$ collisions 
provided by the BEPCII storage ring~\cite{Yu:IPAC2016-TUYA01}
in the center-of-mass energy range from 2.0 to 4.95~GeV,
with a peak luminosity of $1 \times 10^{33}\;\text{cm}^{-2}\text{s}^{-1}$ 
achieved at $\sqrt{s} = 3.77\;\text{GeV}$. 
BESIII has collected large data samples in this energy region~\cite{Ablikim:2019hff, EcmsMea, EventFilter}. The cylindrical core of the BESIII detector covers 93\% of the full solid angle and consists of a helium-based
 multilayer drift chamber~(MDC), a plastic scintillator time-of-flight
system~(TOF), and a CsI(Tl) electromagnetic calorimeter~(EMC),
which are all enclosed in a superconducting solenoidal magnet
providing a 1.0~T magnetic field.
The solenoid is supported by an
octagonal flux-return yoke with resistive plate counter muon
identification modules interleaved with steel. 
The charged-particle momentum resolution at $1~{\rm GeV}/c$ is
$0.5\%$, and the 
${\rm d}E/{\rm d}x$
resolution is $6\%$ for electrons
from Bhabha scattering. The EMC measures photon energies with a
resolution of $2.5\%$ ($5\%$) at $1$~GeV in the barrel (end cap)
region. The time resolution in the TOF barrel region is 68~ps, while
that in the end cap region was 110~ps. The end cap TOF
system was upgraded in 2015 using multigap resistive plate chamber
technology, providing a time resolution of
60~ps,
which benefits 86\% of the data used in this analysis~\cite{etof}.

Simulated data samples produced with a {\sc
geant4}-based~\cite{geant4} Monte Carlo (MC) package, which
includes the geometric description of the BESIII detector and the
detector response, are used to optimize event selection criteria, determine detection efficiencies and estimate backgrounds. The simulation models the beam
energy spread and initial state radiation (ISR) in the $e^+e^-$
annihilations with the generator {\sc
kkmc}~\cite{ref:kkmc}. The inclusive MC sample includes the production of the
$\psi(3686)$ resonance, the ISR production of the $J/\psi$ resonance, and
the continuum processes incorporated in {\sc
kkmc}. All particle decays are modelled with {\sc
evtgen}~\cite{ref:evtgen} using branching fractions 
either taken from the
Particle Data Group (PDG)~\cite{pdg}, whenever available,
or otherwise estimated with {\sc lundcharm}~\cite{ref:lundcharm}. In this analysis, we also use the BODY3~\cite{BODY3} model to generate signal MC events with intermediate structures taken into consideration.


\section{Event Selection}\label{sec:selection}

The $\Lambda(\bar{\Lambda})$ candidates are reconstructed via $\Lambda(\bar{\Lambda}) \to p \pi^{-}(\bar{p}\pi^{+})$, and the $\omega$ candidate is reconstructed via $\omega\to \pi^{+}\pi^{-}\pi^{0}$. 
We also detect the radiative photon from the decay $\psi(3686)\to\gamma \chi_{cJ}$.  Thus all final state particles are reconstructed in the chain $\psi(3686)\to\gamma \chi_{cJ}$ with $\chi_{cJ} \to \Lambda\bar{\Lambda}\omega$.

Candidate events must contain at least three positively charged tracks and three negatively charged tracks. Furthermore, the polar angle of each track measured in the MDC is required to satisfy ${\left|\cos\theta\right|} < 0.93$, where $\theta$ denotes the polar angle defined with respect to the $z$ axis, which is the symmetry axis of the MDC. The d$E/$dx information in the MDC, together with the time of flight in the TOF detector are combined to identify the type of particle (PID). For this purpose, confidence levels for pion, proton and kaon hypotheses are calculated and tracks are assigned to the hypothesis with the highest confidence level. Since the $\Lambda(\bar{\Lambda})$ has a relatively long lifetime, we require the decay length of the $\Lambda(\bar{\Lambda})$ to be greater than zero, where the decay length represents the distance between the interaction point (IP) to the decay position of the $\Lambda(\bar{\Lambda})$. 
The $\Lambda$ and $\bar{\Lambda}$ candidates are reconstructed by combining pairs of oppositely charged tracks with pion and proton mass hypotheses, fulfilling a secondary vertex constraint~\cite{vertex_second}. Only the best combination with the smallest $\Delta M = \sqrt{ (M(p\pi^-)-m(\Lambda))^2+(M(\bar p \pi^+)-m(\bar\Lambda))^2 }$ is retained, where $M(p\pi^-)(M(\bar p \pi^+))$ is the invariant mass of the $p\pi^-(\bar p \pi^+)$ system, and $m(\Lambda)(m(\bar \Lambda))$ is the nominal mass of the $\Lambda(\bar{\Lambda})$~\cite{pdg}.
The $\pi^{+}$ and $\pi^{-}$ candidates from the $\omega$ are selected from the set of pions not assigned to the $\Lambda(\bar{\Lambda})$. Furthermore, the distance of closest approach to the IP must be less than 10\,cm along the $z$-axis, and less than 1\,cm in the transverse plane.

Good photons are selected using clusters in the EMC with the following requirements. (1) In the barrel region of the EMC $\left( {\left| {\cos \theta } \right|  < 0.80} \right)$, the deposited energy must be greater than 25 MeV, while in the end cap regions $(0.86 <  {\left| {\cos \theta } \right| < 0.92} )$, the deposited energy must be greater than 50 MeV. (2) In order to suppress electronic noise, beam related background and cosmic rays, the difference between the EMC time and the event start time is required to be less than 700 ns. (3)~The total number of photons is required to be at least three.

In order to further suppress background and improve the resolution, a five-constraint (5C) kinematic fit (four constraints for the four momentum and one for the $\pi^0$ mass) is applied to all combinations of the final state candidates.  Only the combination having the minimal $\chi_{5\rm C}^{2}$ is retained for further analysis. A figure of merit (FOM) optimization is performed for the selection on $\chi_{5\rm C}^{2}$, which is based on maximizing the $S\over{\sqrt{S+B}}$ value ($S$ denotes the signal yield from the signal MC sample, $B$ denotes the background yield from the inclusive MC sample normalized to the integrated luminosity of data($\approx$ 3208.53~$pb^{-1}$)). The optimized $\chi_{5\rm C}^{2}$ selection criterion is found to be 30.

The combinatorial backgrounds, mainly composed of $\chi_{cJ}\to\Sigma^{*}\bar{\Sigma}^{*}\pi$, are vetoed by imposing additional selection criteria, as listed in Table~\ref{list:vetoes}. The resolution has been subtracted in the selection criteria. The multiple photon background channels are suppressed by requiring $\chi_{\rm signal}^2 < \chi_{\rm bkg}^2$, where $\chi_{\rm signal}^2$ denotes the $\chi^{2}$ under the hypothesis of $\chi_{cJ}\to\Lambda\bar{\Lambda}\omega$, while $\chi_{\rm bkg}^{2}$ denotes the $\chi^{2}$
under the hypothesis of $\psi(3686)\to\pi^{+}\pi^{-}J/\psi$ or $\psi(3686)\to\Lambda\bar{\Lambda}\omega$.

To veto $\chi_{cJ}\to\Sigma^{*-}\Sigma^{*0}\pi^{+}$ and their charge conjugate channels, we require $M(\bar{\Lambda}\pi^{0})$ and $M(\Lambda\pi^{-})$ to be outside the $\bar{\Sigma}^{*0}$ and $\Sigma^{*-}$ signal windows. The same criteria are imposed on their charge conjugate channel. To veto $\chi_{cJ}\to\bar{\Xi}^{+}\Xi^{-}\pi^{0}$, we require $M(\bar{\Lambda}\pi^{+})$ and $M(\Lambda\pi^{-})$ to be outside the $\bar{\Xi}^{+}$ and $\Xi^{-}$ signal windows. To veto $\chi_{cJ}\to\Xi^{-}\bar{\Xi}^{0}\pi^{+}$, we require $M(\Lambda\pi^{-})$ and $M(\bar{\Lambda}\pi^{0})$ to be outside the $\Xi^{-}$ and $\bar{\Xi}^{0}$ signal windows, and the same criteria are imposed on their charge conjugate channel. To veto backgrounds containing $J/\psi$, we require the invariant mass $M(\Lambda\bar{\Lambda}\pi^{+}\pi^{-})$ and the recoil mass $RM(\pi^{+}\pi^{-})$ to be outside the $J/\psi$ signal windows. To veto $\psi(3686)\to\omega\Sigma^{0}\bar{\Sigma}^{0}$, we require $M(\Lambda\gamma^{E1})$ and $M(\bar{\Lambda}\gamma^{E1})$ to be outside the $\Sigma^{0}$ and $\bar{\Sigma}^{0}$ signal windows. Here the $\Lambda\gamma^{E1}$ denotes the combination of $\Lambda$ and the photon from $\psi(3686)$ radiative decay.
Other potential background events are investigated by analyzing the $\psi(3686)$ inclusive MC sample, with the TopoAna package~\cite{TopoAna}. Only a few events survive the event selection. Analyzing the specific background MC samples for the survived events mentioned above, which include $\chi_{cJ}\to \Sigma^{*0} \bar{\Sigma}^{*+}\pi^{-}, \chi_{cJ}\to \Sigma^{*-} \bar{\Sigma}^{*0}\pi^{+}, \chi_{cJ}\to \Sigma^{*+} \bar{\Sigma}^{*-}\pi^{0}, \chi_{cJ}\to \Xi^{0}\bar{\Xi}^{+}\pi^{-}, \chi_{cJ}\to \Xi^{-}\bar{\Xi}^{+}\pi^{0}, \chi_{cJ}\to \Xi^{-}\bar{\Xi}^{0}\pi^{+}, \chi_{cJ}\to \rho^{0}\Sigma^{0}\bar{\Sigma}^{0}, \chi_{cJ}\to \rho^{+}\Lambda \bar{\Sigma}^{*-}, \chi_{cJ}\to \Sigma^{0} \bar{\Sigma}^{0}\omega, \chi_{cJ}\to \Sigma^{*0} \bar{\Sigma}^{*-}\pi^{+}, \chi_{cJ}\to \Sigma^{*-} \bar{\Sigma}^{*+}\pi^{0}, \chi_{c0}\to \Sigma^{*+} \bar{\Lambda}\rho^{-}, \chi_{cJ}\to \Sigma^{*+} \bar{\Sigma}^{*0}\pi^{-}$, we find that the normalized background fraction is less than $1\%$ and thus can be safely ignored.
We impose the same event selection criteria for the continuum data taken at $\sqrt{s}=$ 3.650 GeV. After all the selection criteria, only one event survives. Therefore, the continuum contribution is also negligible. 

\begin{figure}[htbp]
\begin{center}
\begin{minipage}[t]{0.415\linewidth}
\includegraphics[width=1.0\textwidth]{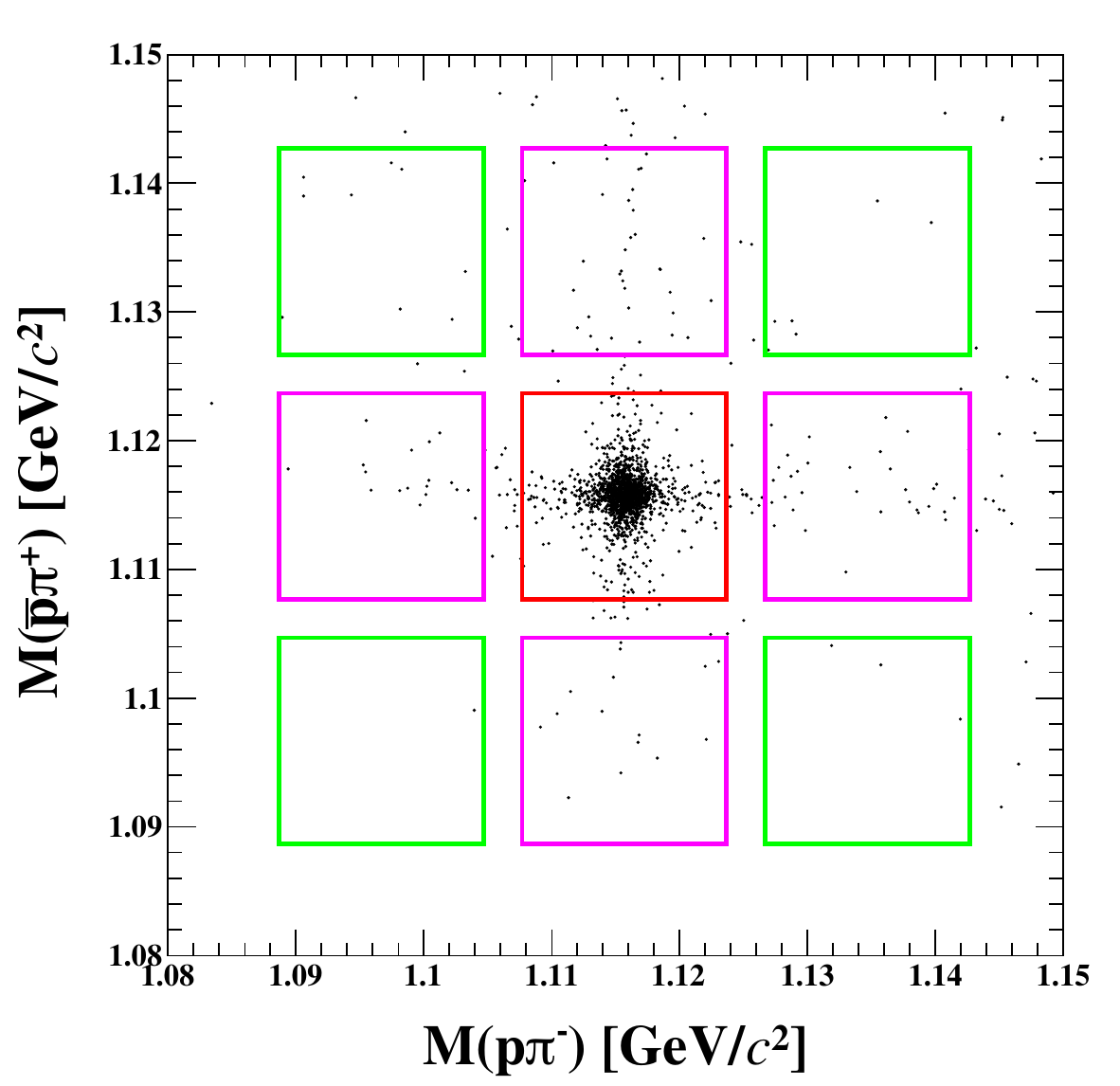}
\end{minipage}%
\begin{minipage}[t]{0.55\linewidth}
\includegraphics[width=1.0\textwidth]{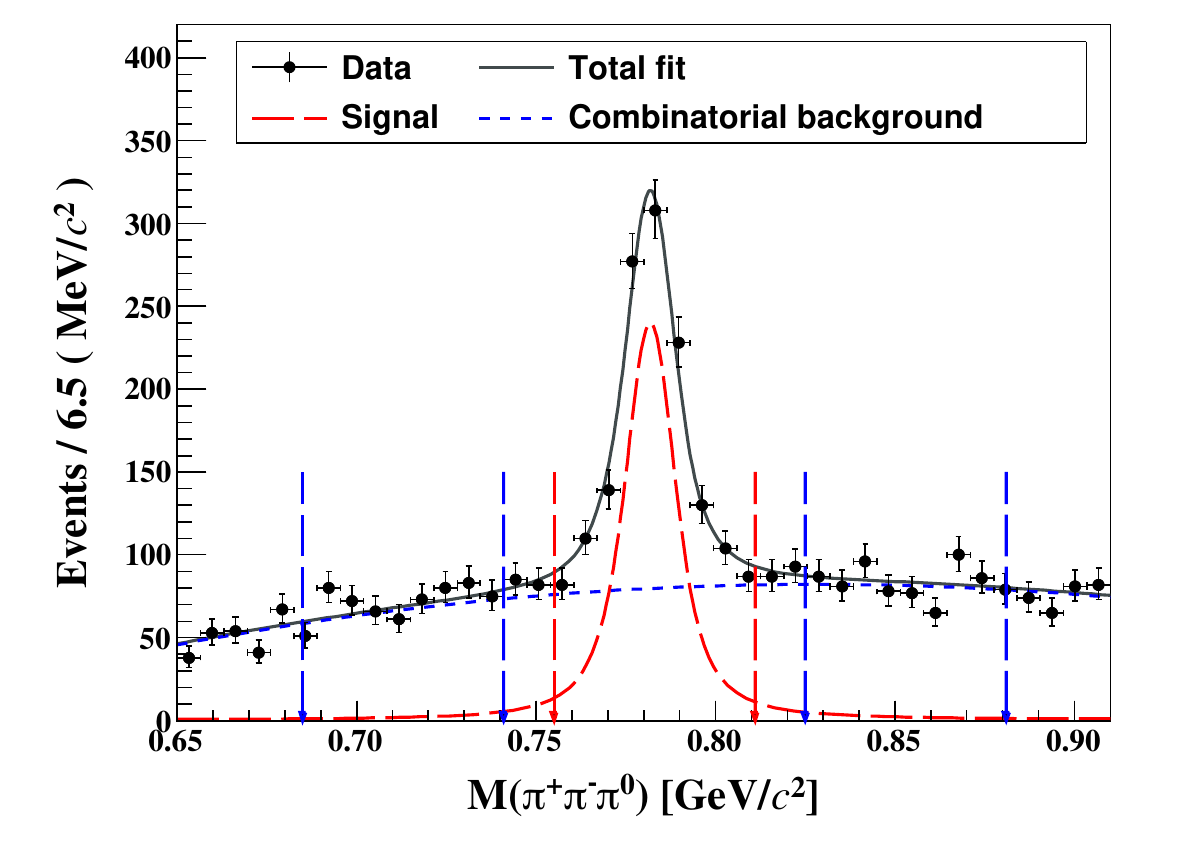}
\end{minipage}%

\caption{The distributions of (left) $M(\bar p \pi^+)$ versus $M(p \pi^-)$ and (right) $M(\pi^{+}\pi^{-}\pi^{0})$ of the accepted candidates. In the left figure, the central red box represents the $\Lambda\bar \Lambda$ signal region, the green boxes are the $\Lambda\bar{\Lambda}$ sideband region 2, and pink boxes are the $\Lambda\bar \Lambda$ sideband region 1. In the right figure, the red dashed line represents the fitted $\omega$ signal, and the blue dashed line denotes the combinatorial background. The grey line is the total fit. The red arrows denote the  $\omega$ signal region, while the blue arrows denote the  $\omega$ sideband regions.}

\label{fig:ConSam_pppipi_scatter_2Lambda}
\end{center}
\end{figure}

The distributions of $M(\bar p \pi^+)$ versus $M(p \pi^-)$ and $M(\pi^{+}\pi^{-}\pi^{0})$ after applying all the selection criteria are illustrated in Fig.~\ref{fig:ConSam_pppipi_scatter_2Lambda}. The $\Lambda(\bar \Lambda)$ signal mass window of $M{(p\pi)}$ is chosen as 8 MeV/$c^{2}$ around the nominal $\Lambda$ mass, and the one dimensional (1D) sideband region is set to be $[1.0887, 1.1047]$ GeV/$c^{2}$ and $[1.1267,1.1427]$~GeV/$c^{2}$. The eight squares with equal areas around the signal region are taken as two dimensional (2D) sideband regions, called $\Lambda\bar \Lambda$ sideband 1 or 2 regions. The $\omega$ signal mass window is taken as $M(\pi^+\pi^-\pi^0)$ $\in$ $[0.756, 0.810]~\rm GeV/c^{2}$.

\begin {table*}[htbp]
\begin{center}
  \fontsize{8}{10}\selectfont

\renewcommand\arraystretch{1.2}

    {\caption {Mass veto windows for different background sources, where m(X) is the nominal mass of the X particle from PDG.}
\label{list:vetoes}}
\begin{tabular}{c|  c}
  \hline \hline

  Veto & Mass window \\   \hline
  \multirow{4}*{\normalsize$\chi_{cJ}\to\Sigma^{*-}\bar{\Sigma}^{*0}\pi^{+}+c.c.$} & $|M(\bar{\Lambda}\pi^{0})-M(\bar{p}\pi^{+})+m(\Lambda)-m(\bar{\Sigma}^{*0})| > 50  $~MeV/$c^{2}$ and \\
   &   $|M(\Lambda\pi^{-})-M(p\pi^{-})+m(\Lambda)-m(\Sigma^{*-})| > 50  $~MeV/$c^{2}$  \\
   
   &  $|M(\bar{\Lambda}\pi^{+})-M(\bar{p}\pi^{+})+m(\Lambda)-m(\bar{\Sigma}^{*+})| > 50  $~MeV/$c^{2}$ and  \\

   &   $|M(\Lambda\pi^{0})-M(p\pi^{-})+m(\Lambda)-m(\Sigma^{*0})| > 50  $~MeV/$c^{2}$ \\  \hline

\multirow{2}*{\normalsize $\chi_{cJ}\to\bar{\Xi}^{+}\Xi^{-}\pi^{0}$} & $|M(\bar{\Lambda}\pi^{+})-M(\bar{p}\pi^{+})+m(\Lambda)-m(\bar{\Xi}^{+})| > 25  $~MeV/$c^{2}$ and \\
    &  $|M(\Lambda\pi^{-})-M(p\pi^{-})+m(\Lambda)-m(\Xi^-)| > 25  $~MeV/$c^{2}$  \\
  \hline   
\multirow{4}*{\normalsize$\chi_{cJ}\to\Xi^{-}\bar{\Xi}^{0}\pi^{+} +c.c.$} &  $|M(\bar{\Lambda}\pi^{+})-M(\bar{p}\pi^{+})+m(\Lambda)-m(\bar{\Xi}^{+})| > 6  $~MeV/$c^{2}$ and \\ &  $|M(\Lambda\pi^{0})-M(p\pi^{-})+m(\Lambda)-m(\Xi^0)| > 4  $~MeV/$c^{2}$ \\  &
$|M(\bar{\Lambda}\pi^{0})-M(\bar{p}\pi^{+})+m(\Lambda)-M(\bar{\Xi^0})| > 9 $~MeV/$c^{2}$ and\\ & $|M(\Lambda\pi^{-})-M(p\pi^{-})+m(\Lambda)-m(\Xi^-)| > 10  $~MeV/$c^{2}$\\ \hline

\multirow{2}*{\normalsize $J/\psi$} &  $\left|M(\Lambda\bar{\Lambda}\pi^{+}\pi^{-})-m(J/\psi)\right| > 40  $~MeV/$c^{2}$ \\ &

$\left|RM(\pi^{+}\pi^{-})-m(J/\psi)\right| > 3  $~MeV/$c^{2}$\\ \hline
\multirow{2}*{\normalsize$\psi(3686)\to\omega\Sigma^{0}\bar{\Sigma}^{0}$} & $\left|M(\Lambda \gamma^{E1})-M(p\pi^{-})+m(\Lambda)-m(\Sigma^{0})\right|>9$~ MeV/$c^{2}$\\ &
$\left|M(\bar{\Lambda} \gamma^{E1})-M(\bar{p}\pi^{+})+m(\Lambda)-m(\bar{\Sigma}^{0})\right| > 10  $~MeV/$c^{2}$\\\hline\hline
  
\end{tabular}
\end{center}
\end{table*}

\section{Signal yield extraction}\label{Sec:sig}

The signal yields of $\chi_{cJ}$ decays are extracted by performing a simultaneous fit to the $M(\Lambda\bar{\Lambda}\pi^{+}\pi^{-}\pi^{0})$ distributions, with $M_{\pi^+\pi^-\pi^0}$ in the $\omega$ signal and sideband regions, and with $M_{p\pi^-}$ and $M_{\bar p\pi^+}$ in the $\Lambda \bar \Lambda$ signal region. This choice is because the number of $\chi_{cJ}$ events in the $\omega$ sideband region is of the same order as in the $\omega$ signal region. The $\omega$ sideband region is chosen as $[0.693, 0.747]~\rm GeV/c^{2}$ and $[0.819, 0.873]~\rm GeV/c^{2}$.

\begin{figure*}[htbp]
  \begin{center}
  \begin{minipage}[t]{1.0 \linewidth}
  \includegraphics[width=1\textwidth]{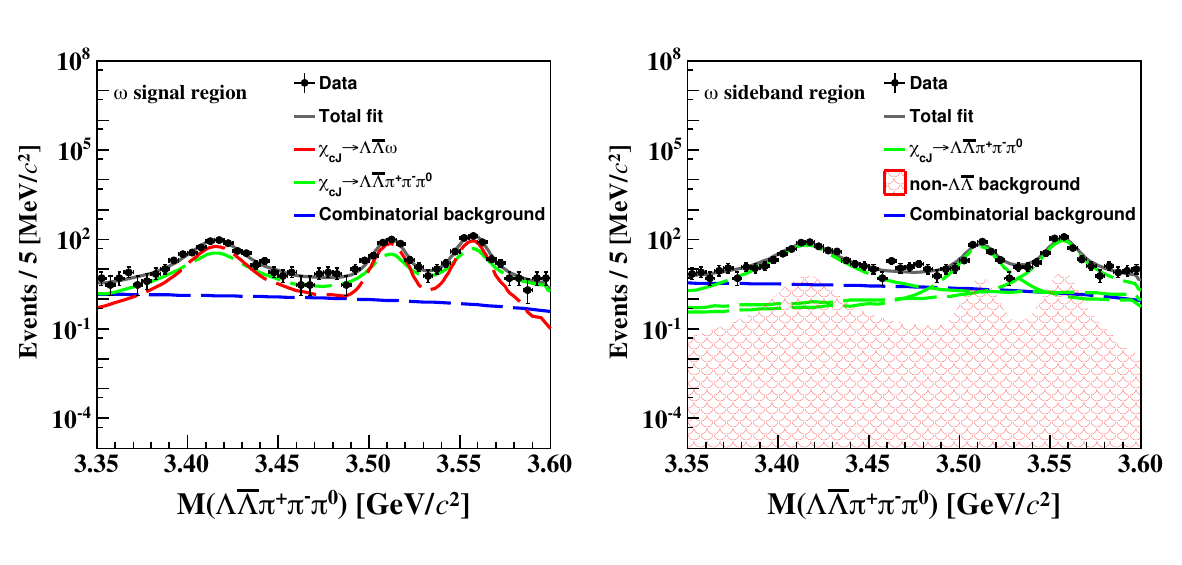}
  \end{minipage}
  \begin{minipage}[t]{1.0 \linewidth}
  \includegraphics[width=1\textwidth]{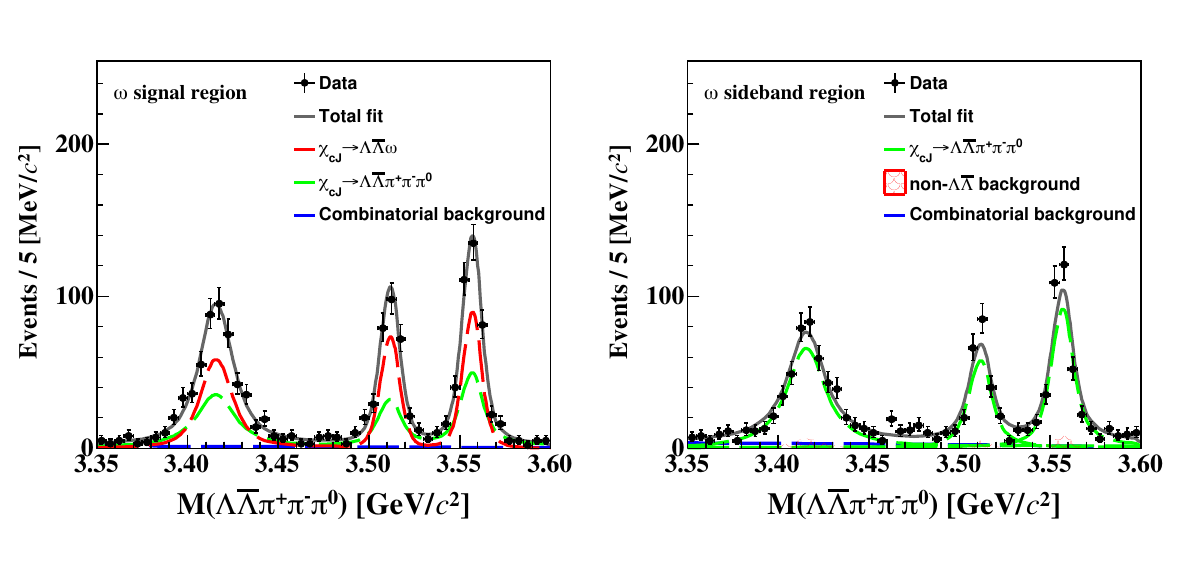}
  \end{minipage}
\caption{ Simultaneous fit to the $M_{\Lambda\bar{\Lambda}\pi^{+}\pi^{-}\pi^{0}}$ distributions in the $\omega$ signal (left) and sideband (right) regions. The first row represents the fit on a logarithmic scale, while the second row shows the normal scale. In the left figures, the red dashed line is the signal, and the green dashed line is the background contribution constrained by the fit to the $\omega$ sideband. Additionally, the blue dashed line is the combinatorial background. The grey line is the total fit. In the right figures, the green line is from the simulated $\chi_{cJ} \to \Lambda\bar{\Lambda} \pi^{+}\pi^{-}\pi^{0}$ shape, and the red histogram is the fixed contribution from the non-$\Lambda\bar{\Lambda}$ background estimated by the $\Lambda\bar \Lambda$ sideband region of data.}
\label{fit_result}
\end{center}
\end{figure*}

For the fit to the events in the $\omega$ signal region, as shown in Fig.~\ref{fit_result}, the probability density functions of the $\chi_{cJ}$ signals are modeled by individual simulated signal shapes convolved with a Gaussian resolution function which describes the resolution difference between data and MC simulation. The combinatorial background is described by a second-order Chebyshev polynomial function(see Fig.~\ref{fit_result} (left)). The non-$\omega$ background is constrained by the simultaneous fit to the events in the $\omega$ sideband region. In the fit to the $\omega$ sideband region, Fig.~\ref{fit_result} (right), the $\chi_{cJ}$ shape convolved with a Gaussian resolution function is taken from the simulated $\chi_{cJ}\to\Lambda\bar{\Lambda}\pi^{+}\pi^{-}\pi^{0}$ channel, and the combinatorial background is described by a second-order Chebyshev polynomial function. To determine the scale factor between the $\omega$ signal and sideband regions, $f_{\omega}$, a fit is performed on the $M(\pi^+\pi^-\pi^0)$ distribution, with results shown in Fig.~\ref{fig:ConSam_pppipi_scatter_2Lambda}(right), in which a simulated signal MC shape convolved with a Gaussian function is used to model the $\omega$ signal and a polynomial function is used to describe the combinatorial background. The $f_{\omega}$ is determined to be 0.530, which is the ratio between the numbers of background events in the $\omega$ signal and sideband regions.

For the above two fits, we have also examined the potential non-$\Lambda\bar \Lambda$ background by using events with one correct $\Lambda(\bar \Lambda)$ and one wrong $\bar \Lambda(\Lambda)$ (marked by the four pink boxes in Fig. 1(left)), and with wrong $\Lambda$ and wrong $\bar \Lambda$ (marked by the four green boxes in Fig. 1(left)).
We find that this kind of background is negligible in the $\omega$ signal region, while it is about $6.3\%$ in the $\omega$ sideband region. This is due to the presence of numerous combinatorial background decays from the $\chi_{cJ}$ that do not include $\Lambda(\bar{\Lambda})$ and $\omega$. Therefore, this kind of background, which is dominated by the $\Lambda\bar \Lambda$ sideband region 1 contribution, is fixed in the fit as shown in Fig.~\ref{fit_result} (right), after normalizing by a factor of 0.5. 

The final fit results are shown in Fig.~\ref{fit_result}. From these fits, we obtain the signal yields for $\chi_{c0,1,2}\to \Lambda\bar \Lambda\omega$ to be $316\pm 30, 202\pm 20$ and $251\pm 23$, with statistical significances of  $11.7\sigma, 11.2\sigma$, and $11.8\sigma$, respectively. For each $\chi_{cJ}$ signal, the statistical significance is calculated with and without including the signal in the fit and considering the change of log-likelihood and taking into account the number of degrees of freedom.

The branching fractions of $\chi_{cJ}\to \Lambda\bar \Lambda\omega$ are calculated by

\begin{equation}
\Br({\chi _{cJ}} \to \Lambda \bar \Lambda \omega ) = {{{N_{\rm fit} \over {N_{\psi(3686)} \cdot \Br\left( {\psi(3686) \to \gamma {\chi _{cJ}}} \right)\cdot {\prod _i}  {\Br_i} \cdot \varepsilon }}}},
\end{equation}
where $N_{\rm fit}$ is the fitted 
  signal yield of $\chicJ$,  ${N_{\psi(3686)}}$ is the total number of $\psi(3686)$ events, ${\mathcal B}(\psi(3686)\to\gamma\chi_{cJ})$ (for $J=0,1,2$) are the branching fractions of $\psi(3696)\to\gamma\chi_{cJ}$, ${\prod _i} {\Br_i}$ is the product of branching fractions of the decays of daughter particles, including $\Br(\Lambda \rightarrow p \pi^{-})=(63.9\pm0.5)\%$, $\Br(\bar{\Lambda} \rightarrow \bar{p} \pi^{+})=(63.9\pm0.5)\%$, $\Br(\pi^{0} \rightarrow \gamma \gamma)=(98.823 \pm 0.034)\%$ and $\Br(\omega \rightarrow \pi^{+} \pi^{-}\pi^{0})=(89.2\pm0.7)\%$ which are taken from the PDG, while $\epsilon$ denotes the detection efficiencies. The details of the fitted signal yields, detection efficiencies, statistical significances, and obtained branching fractions are shown in Table~\ref{list:Br}.




  
   

  
\begin {table}[htbp]
\begin{center}
\fontsize{8}{10}\selectfont

\renewcommand\arraystretch{1.2}

{\caption {Fitted signal yields~($N_{\rm fit}$), 
detection efficiencies ($\epsilon$), statistical significance, and the obtained branching fractions ($\mathcal B$). The first and second uncertainties are statistical and systematic, respectively. 
}
\label{list:Br}}
\begin{tabular}{c|c|c|c|c}
 \hline \hline
 Decay   & $N_{\rm fit}$  & Significance &  $\epsilon~(\%)$ & $\mathcal B~(10^{-4})$ \\   \hline
  
  $\chi_{c 0} \rightarrow \Lambda\bar{\Lambda} \omega$  & $316\pm 30$   & $11.7\sigma$ & 1.38  & $2.37 \pm 0.22 \pm {{0.25}}$ \\
   
 $\chi_{c 1} \rightarrow \Lambda\bar{\Lambda} \omega$  & $202\pm 20$   & $11.2\sigma$ & 2.09  &  $1.01 \pm 0.10 \pm {{0.11}}$ \\

 $\chi_{c 2} \rightarrow \Lambda\bar{\Lambda} \omega$  & $251\pm 23$  & $11.8\sigma$ & 1.92  &   ${{1.40 \pm 0.13 \pm 0.17}}$ \\  \hline\hline

\end{tabular}
\end{center}
\end{table}

\begin{figure*}[htbp]
\begin{center}
\begin{minipage}[t]{0.49\linewidth}
\includegraphics[width=1\textwidth]{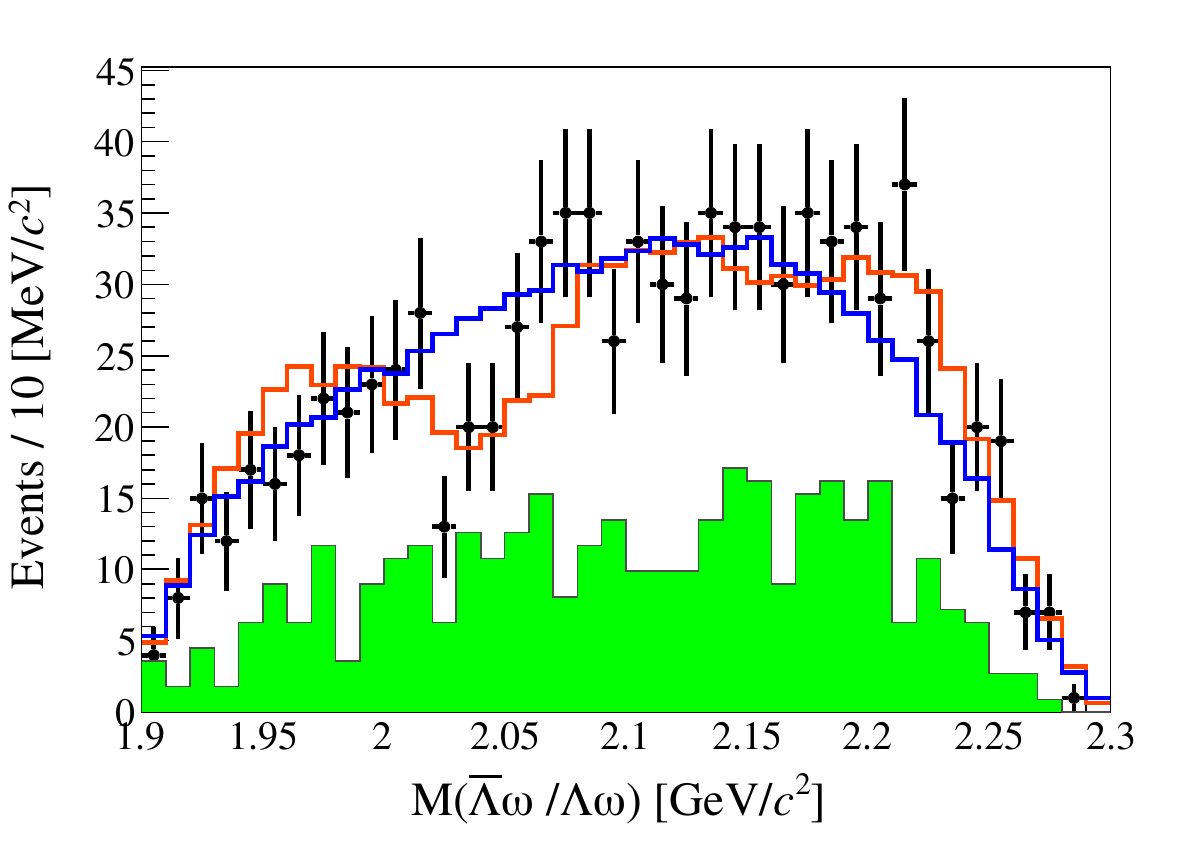}
\end{minipage}
\begin{minipage}[t]{0.49\linewidth}
\includegraphics[width=1\textwidth]{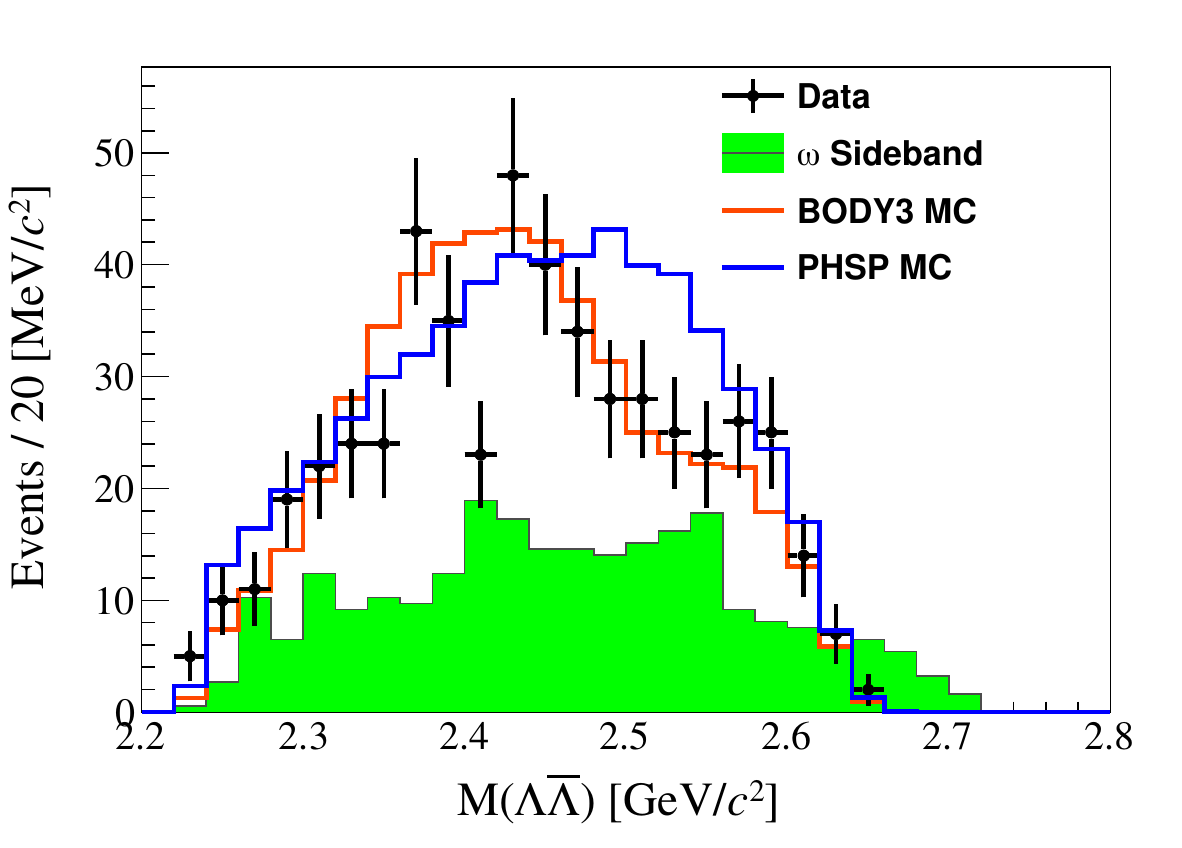}
\end{minipage}

\begin{minipage}[t]{0.49\linewidth}
\includegraphics[width=1\textwidth]{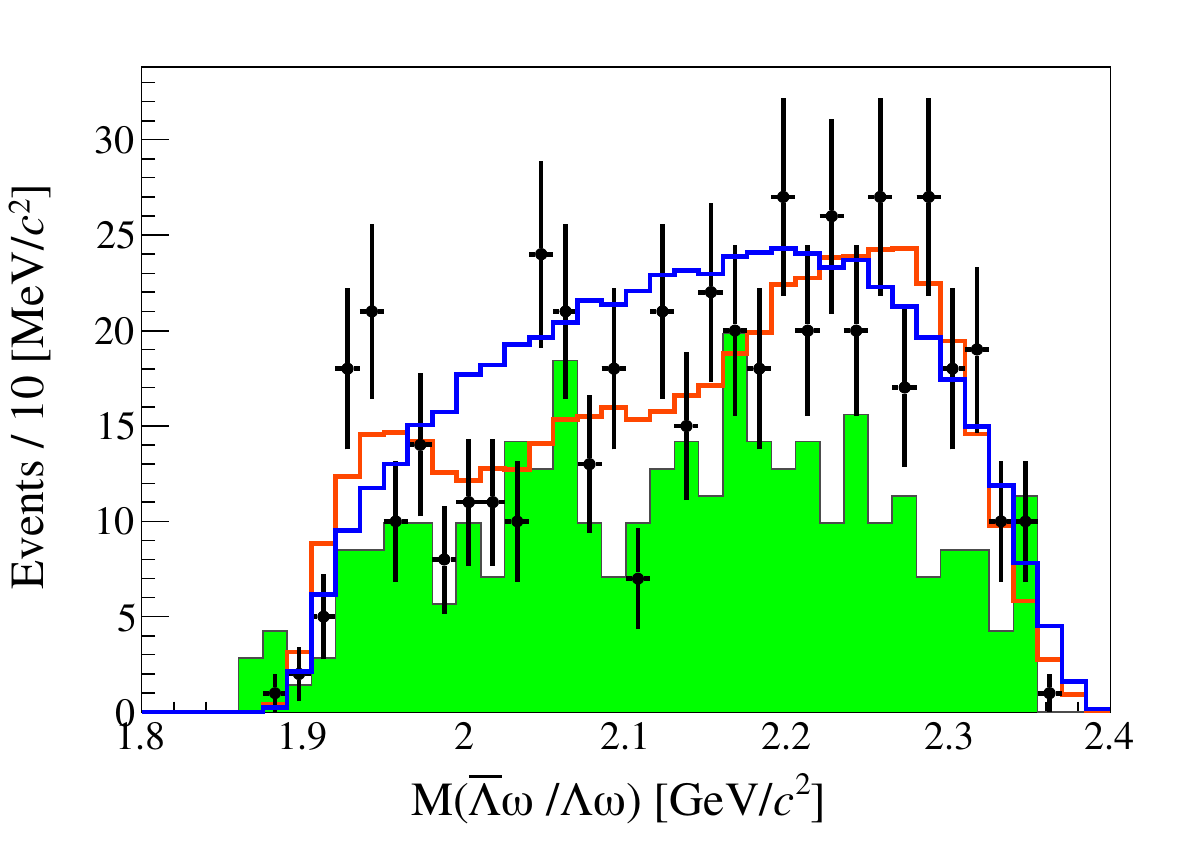}
\end{minipage}
\begin{minipage}[t]{0.49\linewidth}
\includegraphics[width=1\textwidth]{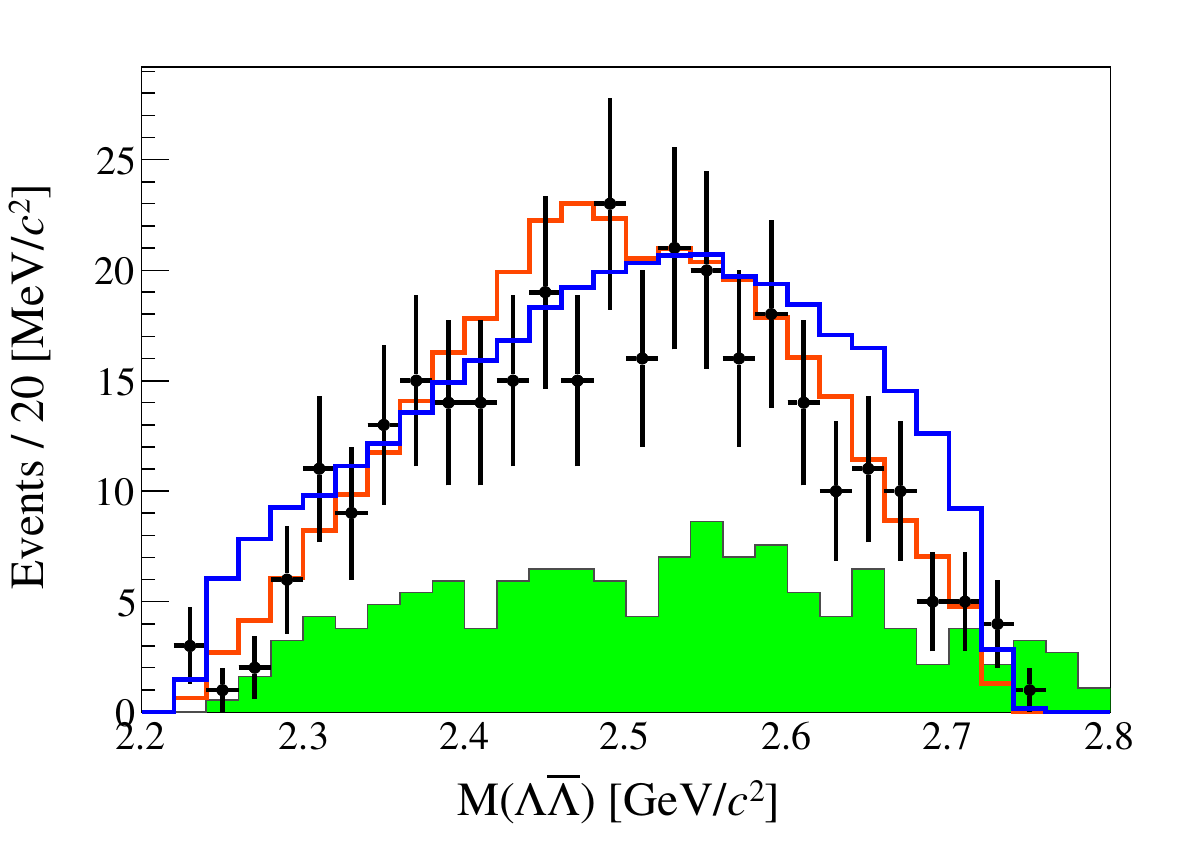}
\end{minipage}

\begin{minipage}[t]{0.49\linewidth}
\includegraphics[width=1\textwidth]{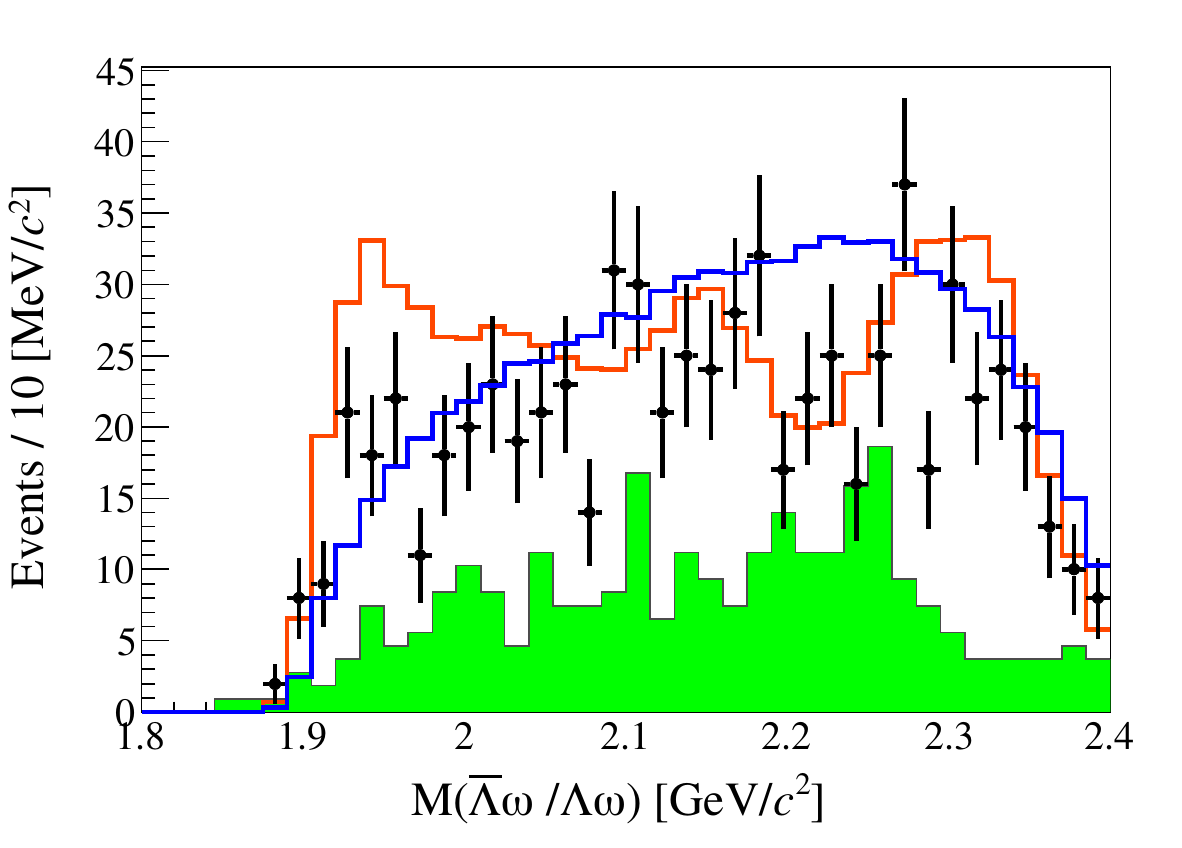}
\end{minipage}
\begin{minipage}[t]{0.49\linewidth}
\includegraphics[width=1\textwidth]{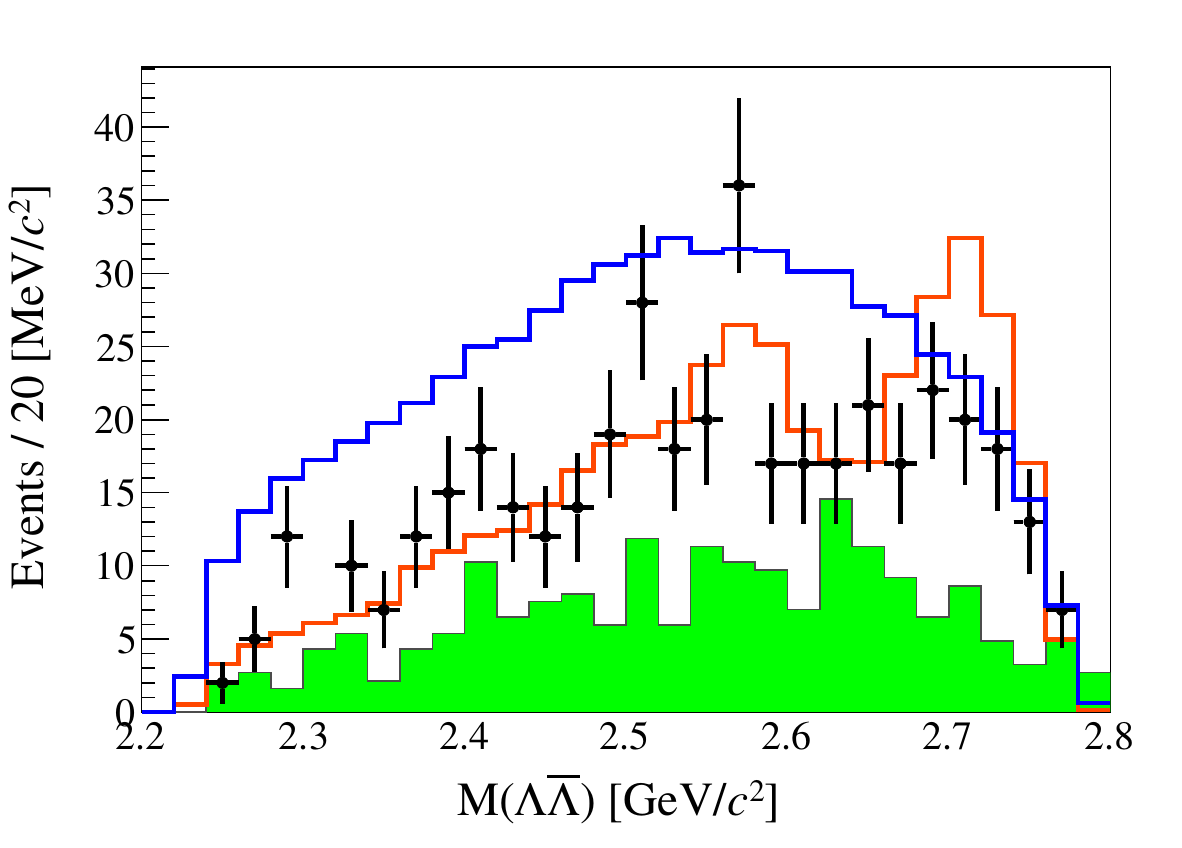}
\end{minipage}

\caption{Comparisons of $M(\bar{\Lambda}\omega /\Lambda\omega)$ and $M(\Lambda\bar{\Lambda})$ of (top) $\chi_{c0}$, (middle) $\chi_{c1}$, (bottom) $\chi_{c2}$, between the data and individual BODY3 signal MC samples after all event selection criteria have been applied. }
\label{BODY3_hists}
\end{center}
\end{figure*}

To investigate possible intermediate structures, we examine the $M(\Lambda\bar{\Lambda})$, $M(\Lambda\omega)$, and $M(\bar{\Lambda}\omega)$ mass distributions from data with background subtracted. No obvious structures are observed with the current statistics. To take into account the slight differences between data and the PHSP signal MC sample, we develop a data-driven BODY3 model. The Dalitz plot of $M^{2}_{\Lambda\omega}$ versus $M^{2}_{\bar \Lambda\omega}$ obtained from data is taken as input for the BODY3 model, which is corrected for backgrounds and efficiencies. The data-MC comparison is shown in Fig.~\ref{BODY3_hists}. 

\section{Systematic Uncertainties}\label{sec:sysU}

The relative systematic uncertainties are from the following sources: tracking and PID efficiencies; the reconstruction efficiency of photons; the reconstruction efficiency of the $\Lambda (\bar{\Lambda})$; the wrong combination background (WCB) of the $\Lambda (\bar{\Lambda})$; the 5C kinematic fit; the mass window selection; the fitting method; the scale factor between the $\omega$ signal and sideband regions; modeling the intermediate states; the quoted branching fractions; and the total number of $\psi(3686)$ events. They are discussed below. 

\begin{itemize}

\item{{\bf Tracking efficiency}}:
The systematic uncertainty due to pion tracking is estimated using the control sample of $J/\psi  \to K_S^0{K^ \pm }{\pi ^ \mp }$. It is estimated to be $1.0\%$ per pion \cite{Ablikim:2011kv}.

\item{\bf PID for pion:} The systematic uncertainty due to the PID efficiency is estimated to be $1.0\%$ for each pion~\cite{Ablikim:2011kv}. We only use PID for the two pions from the $\omega$. Thus the systematic uncertainty due to pion PID is assigned as $2.0\%$.

\item{{\bf Photon reconstruction}}:
The systematic uncertainty due to photon reconstruction is estimated with the control sample of $J/\psi \to \pi^{+}\pi^{-}\pi^{0}$, and is estimated to be $0.5\%$ for each photon \cite{Ablikim:2010zn}. There are three photons in the final states, so the total systematic uncertainty due to photon reconstruction is assigned as $1.5\%$.

\item{\bf $\Lambda/\bar \Lambda$ reconstruction}:
The reconstruction efficiency of $\Lambda (\bar{\Lambda})$ including tracking and reconstruction  is estimated by using the control sample of $J/\psi\to\bar{\Lambda}\pi^{+}\Sigma^{-}+c.c.$ \cite{yipuLiao}, in which the signal MC is reweighted within each bin of $p$ and $|cos \theta|$, where the p denotes the momentum of $\Lambda$. So the uncertainties are different due to the different $p$ and $|cos \theta|$ distributions of $\Lambda$ decays from $\chi_{c0, 1, 2}$. The systematic uncertainties of $\Lambda$ reconstruction are assigned as 2.5\%, 2.7\% and 2.9\% for $\chi_{c0,1,2}$, and those of $\bar \Lambda$ reconstruction are assigned as 2.2\%, 2.4\% and 2.7\% for $\chi_{c0,1,2}$, respectively. 

\item{\bf WCB of $\Lambda(\bar \Lambda)$}:
The potential impact of WCB in the reconstruction of $\Lambda (\bar{\Lambda})$ is investigated by matching the angles of the MC generated and reconstructed track momenta for the $\Lambda$. Events with angle differences greater than $3^{\circ}$ are considered as mismatched backgrounds. Any possible bias related to WCB is examined by comparing the signal yields with and without the WCB component included in the fit. The differences on the branching fractions are taken as the systematic uncertainties, which are 1.3\%, 0.7\% and 0.4\% for $\chi_{c0, 1, 2}\to\Lambda\bar{\Lambda}\omega$, respectively.

\item{\bf 5C kinematic fit}:
The systematic uncertainty associated with the 5C kinematic fit is assigned as the difference between the efficiencies before and after the helix correction \cite{HelixG}, which are 3.1\%, 3.2\% and 2.9\% for $\chi_{c0, 1, 2}\to\Lambda\bar{\Lambda}\omega$, respectively.

\item{\bf Mass window}:
    To estimate the systematic uncertainties due to the mass windows of $\Lambda, \bar{\Lambda}$ and $\omega$, we use the smearing method. From the simultaneous fit, the parameters of the Gaussian function which is used to compensate the resolution difference between data and MC simulation are obtained. Then we smear the simulated signal with the Gaussian function with these obtained parameters, and the difference between the efficiencies before and after smearing the Gaussian resolution function is taken as the systematic uncertainty. For the mass windows of $\Lambda$ and $\bar \Lambda$, the systematic uncertainties are negligible for all signal decays. For the mass window of $\omega$, the systematic uncertainties are 0.1\%, 0.8\% for $\chi_{c0,1}\to\Lambda\bar{\Lambda}\omega$, respectively; and that of $\chi_{c2}\to\Lambda\bar{\Lambda}\omega$ is negligible.
\item{\bf Mass vetoes}:
    To estimate the systematic uncertainties for each mass veto, we examine the branching fractions after enlarging or shrinking the veto region. For different background vetoes, we vary the corresponding mass windows seven times with a step of 1, 2, or 4 MeV$/c^2$. For each case, the deviation between the alternative and nominal fits is defined as  $\zeta =$ ${|\mathcal B_{\rm nominal}-\mathcal B_{\rm test}|}\over{\sqrt{|\sigma^{2}_{\mathcal B,\rm nominal}-\sigma^{2}_{\mathcal B,\rm test}|}}$, where ${\mathcal B}$ denotes the branching fractions of $\chi_{cJ}\to\Lambda\bar{\Lambda}\omega$ and $\sigma$ denotes its statistical uncertainty. If $\zeta$ is less than 2.0, the associated systematic uncertainty is negligible according to the Barlow test~\cite{BarlowRef}. Otherwise, its relative difference is assigned as the systematic uncertainty. 

 \item{\bf Fit range}:
   The systematic uncertainties due to the fit range are examined by enlarging and shrinking the fit range seven times, with a 4 MeV$/c^{2}$ change per step, and the Barlow test is performed as above. The systematic uncertainties are negligible for $\chi_{c0,1}$, and 0.9\% is obtained for the $\chi_{c2}$ decay.

 \item{\bf Signal shape}:
The systematic uncertainty arising from the signal shape is evaluated by comparing the fitted results obtained from two distinct simulated signal MC samples. In one sample, the $\Lambda/\bar{\Lambda}$ is generated using the PHSP generator, while in the other, it is generated with the HypWK generator~\cite{HypWK}. The differences in the measured branching fractions are taken as the systematic uncertainties, which are 1.5\%, 5.5\% and 0.2\% for $\chi_{c0, 1, 2}\to\Lambda\bar{\Lambda}\omega$, respectively.

\item{\bf Background shape}:
   The systematic uncertainty due to the background shape is estimated by replacing the second-order Chebyshev polynomial function with a first or third-order Chebyshev polynomial function. The largest differences in the measured branching fractions are taken as the systematic uncertainties, which are 0.8\%, 0.3\% and 0.1\% for $\chi_{c0, 1, 2}\to\Lambda\bar{\Lambda}\omega$, respectively.

\item{\bf Scale factor $f_\omega$}:
The $f_{\omega}$ directly affects the fitted signal yields.  The associated systematic uncertainty is estimated by changing the $\omega$ sideband region by $\pm 1\sigma$, where $\sigma$ denotes the mass resolution of the $\omega$. The largest deviations of the branching fractions are taken as the systematic uncertainties, which are $0.5\%, 2.4\%$ and $1.6\%$ for $\chi_{c0,1,2}\to\Lambda\bar{\Lambda}\omega$, respectively.
    
\item{\bf Background level}:
Additionally, the normalization factor of $f_{\omega}$ also impacts the background level for the BODY3 model. The changes in signal efficiencies are considered as corresponding systematic uncertainties after adjusting the number of events in the sideband region by $\pm1\sigma$, which serves as the input for the BODY3 generator. The systematic uncertainties are assigned as $2.5\%, 0.3\%$ and $1.7\%$ for $\chi_{c0,1,2}\to\Lambda\bar{\Lambda}\omega$, respectively.


\item{\bf BODY3 model}: To estimate the systematic uncertainty associated with the BODY3 model, we also take the binning effect into consideration by changing the number of bins by $\pm25\%$. In each case, the Dalitz plot is obtained by reweighting the PHSP MC sample using the data subtracted background in the $\omega$ and 2D $\Lambda\bar{\Lambda}$ sidebands, and the largest differences of signal efficiencies are taken as the systematic uncertainties, which are $1.6\%, 1.1\%$ and $4.4\%$ for $\chi_{c0, 1, 2}\to\Lambda\bar{\Lambda}\omega$, respectively.
    
\item{\bf Quoted branching fractions}:
The branching fractions of $\psi(3686)\to\gamma\chi_{cJ}, \Lambda\to p \pi^{-}, \bar{\Lambda}\to\bar{p}\pi^{+}, \omega\to\pi^{+}\pi^{-}\pi^{0}$, and $\pi^{0}\to\gamma\gamma$ are quoted from the PDG. Their corresponding uncertainties are taken to calculate the systematic uncertainties, which are $2.7\%, 3.0\%$ and $2.7\%$ for $\chi_{c0, 1, 2}\to\Lambda\bar{\Lambda}\omega$, respectively. 

\item{\bf Total number of $\psi(3686)$ events}:
The uncertainty of the total number of $\psi(3686)$ events, which is
determined with the inclusive hadronic $\psi(3686)$ decays, is assigned as 0.5\%~\cite{Ablikim:2017wyh}.

\end{itemize}

\begin{table}[htp]
\begin{center}
\caption{The relative systematic uncertainties (in percent) in the measurements of the branching fractions of the $\chi_{cJ}\to\Lambda\bar{\Lambda}\omega$ decays. }
\label{list_sys}
\begin{tabular}{l  c  c  c}
\hline\hline  Source & $\chi_{c 0}$ & $\chi_{c 1}$ & $\chi_{c 2}$ \\
\hline   Tracking for $\pi^{\pm}$  & $2.0$  & $2.0$  & $2.0$  \\
 
  $\pi^\pm$ PID   & $2.0$ & $2.0$  & $2.0$  \\
  Photon reconstruction &  $1.5$    &   $1.5$   &  $1.5$   \\
  $\Lambda$ reconstruction  & 2.5  & 2.7  & 2.9  \\
  $\bar \Lambda$ reconstruction  & 2.2  & 2.4  & 2.7  \\
   WCB of $\Lambda(\bar \Lambda)$ & 3.6 & 1.4 & 1.1 \\
   5C kinematic fit &  3.1  &  3.2  &  2.9  \\
    $\Lambda$ signal region & -- & -- & -- \\ 
 $\bar{\Lambda}$ signal region & -- & -- & -- \\
 $\omega$ signal region & 0.1 &0.8 &--\\

 Veto $J/\psi(\Lambda\Lambda\pi^{+}\pi^{-})$  & --  & -- & --  \\
  Veto $J/\psi(\pi^{+}\pi^{-}_{\rm rec})$  & 5.1  & 3.2 & 1.8  \\
  Veto $\gamma\Lambda$ & -- & -- &1.6 \\
  Veto $\gamma\bar{\Lambda}$ & -- & -- & 4.2 \\
  Veto $\Xi^{+}\Xi^{-}$  & --  & 1.2 & --  \\
  Veto $\Xi^{-}\bar{\Xi}^{0}$  & 0.9  & -- & 1.0  \\
  Veto $\bar{\Xi}^{+}\Xi^{0}$  & 2.4  & -- & --  \\
  Veto $\bar{\Sigma}^{+}\Sigma^{0}$ & -- & -- & -- \\
  Veto $\Sigma^{-}\bar{\Sigma}^{0}$  & 1.5  & 2.0 & 6.0  \\ 
  Fit range  & 0.7 & --  & {0.5}  \\

  Signal shape & 1.5 & 5.5  & 0.2  \\
  Background shape & 0.8 & 0.3 & 0.1\\
Scale factor $f_{\omega}$  & {0.5}  & {2.4} & {1.6}  \\


 {{Background level for BODY3 Model}} & 2.5  & 0.3  & 1.7\\

 Binning effect for BODY3 Model & {0.5} & {1.0} & {4.4}\\
 Quoted branching fractions  &  2.7  &  3.0  &  2.7 \\ 

 Total number of $\psi(3686)$ events & {{0.5}}  & {{0.5}}  & {{0.5}}  \\

\hline Total  & {10.4}  & {{10.6}}  & {{12.0}}  \\
\hline\hline

\end{tabular}
\end{center}
\end{table}

All sources of systematic uncertainty and their contributions are summarized in Table~\ref{list_sys}. Under the assumption that the systematic uncertainties of reconstructions and quoted branching fractions for the $\Lambda$ are correlated with those of the $\bar \Lambda$, and all other systematic uncertainties are independent, the total systematic uncertainty for each signal decay is obtained by adding all uncertainties in quadrature, excluding those that are correlated.

	
\section{Summary}\label{sec:summary}
	
The decays $\chi_{cJ}\to\Lambda\bar{\Lambda}\omega$ ($J=0,1,2$) are observed for the first time with  statistical significances of $11.7 \sigma, 11.2 \sigma$ and $11.8 \sigma$, using $(27.12\pm0.14)\times 10^8$ $\psi(3686)$ events collected by the BESIII detector.  Their branching fractions are determined to be $\Br\left(\chi_{c 0} \rightarrow \Lambda\bar{\Lambda} \omega\right)=(2.37 \pm 0.22 \pm {{0.25}})\times 10^{-4}, \Br\left(\chi_{c 1} \rightarrow \Lambda\bar{\Lambda} \omega\right)=(1.01 \pm 0.10 \pm {{0.11}})\times 10^{-4}$ and $\Br\left(\chi_{c 2} \rightarrow \Lambda\bar{\Lambda} \omega\right)=({{1.40 \pm 0.13 \pm 0.17}})\times 10^{-4}$, where the first and second uncertainties are statistical and systematic, respectively. The analysis improves our understanding of the properties of $\chi_{cJ}$ particles.  With the current statistics, no deviations are observed between the data and PHSP MC samples for the $M(\Lambda\bar{\Lambda})$ and $M(\Lambda\omega/\bar{\Lambda}\omega)$ distributions.
The analysis aids in the exploration for potential excited baryon states in the $\Lambda\omega~ (\bar{\Lambda}\omega)$ invariant mass spectra and the search for unidentified structures in the $\Lambda\bar{\Lambda}$ invariant mass spectrum.

	\input{acknowledgement.tex}
	

\end{document}

%% file: def-com.tex
\newcommand{\chicJ}{\chi_{cJ}}

\newcommand{\bfg}{\begin{figure}}
\newcommand{\efg}{\end{figure}}
\newcommand{\bitm}{\begin{itemize}}
\newcommand{\eitm}{\end{itemize}}
\newcommand{\bnum}{\begin{enumerate}}
\newcommand{\enum}{\end{enumerate}}
\newcommand{\btbl}{\begin{table}}
\newcommand{\etbl}{\end{table}}
\newcommand{\btbu}{\begin{tabular}}
\newcommand{\etbu}{\end{tabular}}
\newcommand{\bcl}{\begin{center}}
\newcommand{\ecl}{\end{center}}

\newcommand{\beq}{\begin{equation}}
\newcommand{\eeq}{\end{equation}}
\newcommand{\beqr}{\begin{eqnarray}}
\newcommand{\eeqr}{\end{eqnarray}}

\definecolor{boslv}{rgb}{0.0, 0.65, 0.58}
\definecolor{Munsell}{HTML}{00A877}

\newcommand{\Br}{\mathcal{B}}

%% file: authorlist_2023-12-11.tex
\author{
\begin{small}
\begin{center}
M.~Ablikim$^{1}$, M.~N.~Achasov$^{4,c}$, P.~Adlarson$^{75}$, O.~Afedulidis$^{3}$, X.~C.~Ai$^{80}$, R.~Aliberti$^{35}$, A.~Amoroso$^{74A,74C}$, Q.~An$^{71,58,a}$, Y.~Bai$^{57}$, O.~Bakina$^{36}$
% ...
\\
\vspace{0.2cm}
(BESIII Collaboration)\\
\vspace{0.2cm} {\it
$^{1}$ Institute of High Energy Physics, Beijing 100049, People's Republic of China\\
% ...
$^{3}$ Bochum  Ruhr-University, D-44780 Bochum, Germany\\
$^{4}$ Budker Institute of Nuclear Physics SB RAS (BINP), Novosibirsk 630090, Russia\\
% ...
$^{35}$ Johannes Gutenberg University of Mainz, Johann-Joachim-Becher-Weg 45, D-55099 Mainz, Germany\\
$^{36}$ Joint Institute for Nuclear Research, 141980 Dubna, Moscow region, Russia\\
% ...
$^{57}$ Southeast University, Nanjing 211100, People's Republic of China\\
$^{58}$ State Key Laboratory of Particle Detection and Electronics, Beijing 100049, Hefei 230026, People's Republic of China\\
% ...
$^{71}$ University of Science and Technology of China, Hefei 230026, People's Republic of China\\
% ...
$^{74}$ University of Turin and INFN, (A)University of Turin, I-10125, Turin, Italy; (B)University of Eastern Piedmont, I-15121, Alessandria, Italy; (C)INFN, I-10125, Turin, Italy\\
$^{75}$ Uppsala University, Box 516, SE-75120 Uppsala, Sweden\\
% ...
$^{80}$ Zhengzhou University, Zhengzhou 450001, People's Republic of China\\
\vspace{0.2cm}
$^{a}$ Deceased\\
% ...
$^{c}$ Also at the Novosibirsk State University, Novosibirsk, 630090, Russia\\
% ...
}
\end{center}
\vspace{0.4cm}
\end{small}
}

%% file: acknowledgement.tex
\section{Acknowledgement}

The BESIII Collaboration thanks the staff of BEPCII and the IHEP computing center for their strong support. This work is supported in part by National Key R\&D Program of China under Contracts Nos. 2020YFA0406300, 2020YFA0406400; National Natural Science Foundation of China (NSFC) under Contracts Nos. 11635010, 11735014, 11835012, 11935015, 11935016, 11935018, 11961141012, 12025502, 12035009, 12035013, 12061131003, 12192260, 12192261, 12192262, 12192263, 12192264, 12192265, 12221005, 12225509, 12235017, 12150004; Program of Science and Technology Development Plan of Jilin Province of China under Contract No.20210508047RQ and 20230101021JC; the Chinese Academy of Sciences (CAS) Large-Scale Scientific Facility Program; the CAS Center for Excellence in Particle Physics (CCEPP); Joint Large-Scale Scientific Facility Funds of the NSFC and CAS under Contract No. U1832207; CAS Key Research Program of Frontier Sciences under Contracts Nos. QYZDJ-SSW-SLH003, QYZDJ-SSW-SLH040; 100 Talents Program of CAS; The Institute of Nuclear and Particle Physics (INPAC) and Shanghai Key Laboratory for Particle Physics and Cosmology; European Union's Horizon 2020 research and innovation programme under Marie Sklodowska-Curie grant agreement under Contract No. 894790; German Research Foundation DFG under Contracts Nos. 455635585, Collaborative Research Center CRC 1044, FOR5327, GRK 2149; Istituto Nazionale di Fisica Nucleare, Italy; Ministry of Development of Turkey under Contract No. DPT2006K-120470; National Research Foundation of Korea under Contract No. NRF-2022R1A2C1092335; National Science and Technology fund of Mongolia; National Science Research and Innovation Fund (NSRF) via the Program Management Unit for Human Resources \& Institutional Development, Research and Innovation of Thailand under Contract No. B16F640076; Polish National Science Centre under Contract No. 2019/35/O/ST2/02907; The Swedish Research Council; U. S. Department of Energy under Contract No. DE-FG02-05ER41374.
